\documentclass[prb,twocolumn,amsmath,amssymb,floatfix,superscriptaddress]{revtex4-1}
\usepackage{graphicx}
\usepackage[FIGTOPCAP]{subfigure}
\usepackage[usenames]{color}
\newcommand{\be}{\begin{eqnarray}}
\newcommand{\ee}{\end{eqnarray}}

\newcommand{\wbe}{\begin{widetext}}
\newcommand{\wee}{\end{widetext}}

\begin{document}
\title{Effects of surface-bulk hybridization in 3D topological `metals'}
\date{\today}
\author{Yi-Ting Hsu}
\affiliation{Department of Physics, Cornell University, Ithaca, New York 14853, USA}
\author{Mark H. Fischer}
\affiliation{Department of Physics, Cornell University, Ithaca, New York 14853, USA}
\affiliation{Department of Condensed Matter Physics, Weizmann Institute of Science, Rehovot 76100, Israel}
\author{Taylor L. Hughes}
\affiliation{Department of Physics, University of Illinois, 1110 West Green St, Urbana IL 61801, USA}

\author{Kyungwha Park}
\affiliation{Department of Physics, Virginia Tech, Blacksburg, Virginia 24061, USA}
\author{Eun-Ah Kim}
\affiliation{Department of Physics, Cornell University, Ithaca, New York 14853, USA}

\begin{abstract}
Identifying the effects of surface-bulk coupling is a key challenge in exploiting the topological nature of the surface states in many available three-dimensional topological `metals'. Here we combine an effective-model calculation and an ab-initio slab calculation to study the effects of the lowest order surface-bulk interaction: hybridization. In the effective-model study, we discretize an established low-energy effective four-band model and introduce
hybridization between surface bands and bulk bands in the spirit of the Fano model. We find that hybridization enhances the energy gap between bulk and Dirac surface states and preserves the latter's spin texture qualitatively albeit with a reduced spin-polarization magnitude. Our ab-initio study finds the energy gap between the bulk and the surface states to grow upon an increase in the slab thickness, very much in qualitative agreement with the effective model study. Comparing the results of our two approaches, we deduce that the experimentally observed low magnitude of the spin polarization can be attributed to a hybridization-type surface-bulk interaction. 
We discuss evidence for such hybridization in existing ARPES data.   
\end{abstract}

\maketitle

\section{Introduction}
Many discrepancies between experimental measurements and theoretical predictions of ideal topological insulators (TI's) are attributed to 
the fact that the chemical potential lies in the conduction band and the bulk band interferes with measurements
\cite{PhysRevLett.107.056803,PhysRevB.81.041405,Bianchi2010,PhysRevB.84.233101}. That is, many available TI materials are actually metallic. Recent developments in thin-film experiments\cite{Zhang2010,PhysRevB.84.233101,PhysRevLett.109.116804,PhysRevB.83.165440,PhysRevB.84.073109} further call for studies of surface-bulk electron interaction in films.
However, explicit first-principles calculations on TI films are limited to very thin slabs with thickness less than 10 nm due to the computational cost. Therefore, there is a need for a simple microscopic model which incorporates surface-bulk interactions  that can be used to study how physical properties depend on the film thickness. 

One important question such a model should address is the effect of surface-bulk interaction on the spin-texture. The surface spin-texture is a key physical characteristic of topological surface states in ideal TIs. While low-energy effective theories guided by symmetries predict perfect spin-momentum locking for topological surface states within the bulk gap, spin- and angle-resolved photoemission spectroscopfy (SARPES) data show lower in-plane spin polarization and total spin magnitude\cite{Hsieh2009,PhysRevLett.106.257004,PhysRevLett.105.266806}. On the other hand, an ab-initio calculation on a few-quintuple-layer (QL) slab\cite{PhysRevLett.105.266806} found both the spin polarization and the total spin magnitude to be much smaller. Although reduction in spin polarization and total spin magnitude are to be anticipated at large surface Fermi-momenta where hexagonal warping manifests\cite{Chen2009,PhysRevLett.103.266801,PhysRevLett.105.076802}, little is understood about the reduction observed at small Fermi-momenta and how the surface-bulk interaction affects the spin texture. However, such understanding is crucial for pursuing technical applications of spin-momentum locking in thin films of topological insulators in the metallic regime\cite{garate:2010,yokoyama:2010,fischer:2013}. 

Our starting point is the observation made by \textcite{PhysRevB.82.195417} that the lowest order electron-electron interaction term between the surface state and the bulk states can be viewed as a hybridization term in the Fano model\cite{Mahan}.
The key effect of hybridization in this low-energy effective theory is to 
spectroscopically separate 
surface states localized on the surface 
from the extended metallic bands. Building on the principles underlying this low-energy effective theory, we study the hybridization effects with a microscopic model of 3D time-reversal invariant strong topological insulators to address the thickness dependence of physical quantities and connect the results to ab-initio slab calculations. 
Specifically, we study TI slabs with finite thickness in the presence of surface-bulk interaction from two complementary perspectives: a simple microscopic model including the lowest order surface-bulk interaction and an ab-initio calculation of a few-QL Bi$_2$Se$_3$. We focus on how the spectroscopic properties and the spin texture evolve as a function of film thickness.

The paper is structured as follows. In section \ref{sec:hyb} we construct a  lattice model  for a slab with surface-bulk hybridization (S-B hybridization) and study the spectroscopic properties of the model as well as the effects of the S-B hybridization on the spin texture. In section \ref{sec:DFT} we present an ab-initio study on 4,5,6-QL Bi$_2$Se$_3$ using density functional theory (DFT) and discuss the insight the simple hybridization model offers in understanding the ab-initio results. We then conclude in section \ref{sec:conclusion} with discussions of implications of our results and open questions. 

\section{Lattice model for a slab with s-b hybridization}
\label{sec:hyb}
\subsection{The Model}
In order to introduce surface-bulk hybridization as a perturbation in the spirit of the Fano model\cite{PhysRevB.82.195417} and study its effects on a slab with finite thickness, we first need a lattice model for a slab. For this, we discretize\footnote{A similar discretization was used in other papers, e.g., G. Rosenberg and M. Franz, Phys. Rev. B {\bf 85}, 195119 (2012).}  the effective continuum model by \textcite{Zhang2009}, which is a four band $\mathbf{k}\cdot\mathbf{p}$ Hamiltonian guided by symmetries and first-principle-calculation results. We then make the system size finite along the vertical axis, i.e., the film growth direction.

The low-energy effective four-band model in Ref.~\onlinecite{Zhang2009} describes a strong 3D TI with rhombohedral crystal structure such as Bi$_2$Se$_3$.  
In this effective model each QL is treated as a layer since the inter-QL coupling is weak, 
and the four lowest-lying spin-orbital bands come from the mixing of the two $P_z$ atomic orbitals from Bi and Se referred to as $P_1$ and $P_2$ and the two spins $\uparrow$, $\downarrow$. We take this effective model written in terms of $4\times 4$ $\Gamma$-matrices and discretize it following the discretization scheme used for 2D TIs in Ref.~\onlinecite{JPSJ.77.031007} such that  the lattice Hamiltonian reduces to the low-energy effective $\mathbf{k}\cdot\mathbf{p}$ Hamiltonian in Ref.~\onlinecite{Zhang2009} in the limit $|\mathbf{k}|$ $\rightarrow 0$. The resulting lattice model in the rotated spin-orbital basis $\{|P_1,\uparrow\rangle$, $-i|P_2,\uparrow\rangle$, $|P_1,\downarrow\rangle$, $i|P_2,\downarrow\rangle\}$ is 
\begin{equation}
H_0=\sum_{\mathbf{k_{\parallel}},k_z}{h_0}(\mathbf{k_{\parallel}},k_z)c^{\dagger}_{\mathbf{k_{\parallel}},k_z}c_{\mathbf{k_{\parallel}},k_z},
\label{eq:H0}
\end{equation}
where $c_{\mathbf{k_{\parallel}}, k_z}^\dagger$ is an operator creating a four spinor in the rotated spin-orbital basis with an in-plane momentum $\mathbf{k_{\parallel}}=(k_x,k_y)$ and perpendicular momentum $k_z$, and
the lattice Hamiltonian\footnote{The lattice Hamiltonian $h_0(\mathbf{k_{\parallel}},k_z)$ does not have in-plane three-fold rotational symmetry of the continuum Hamiltonian. However, we do not expect this to change any of our conclusions in a qualitative manner.} is
\begin{align}
\label{eq:H0lattice}
&{h_0}(\mathbf{k_{\parallel}},k_z)=\epsilon_1(\mathbf{k_{\parallel}},k_z)\mathbb{I}_{4\times 4}+\frac{A_1}{a_z}\sin(k_za_z)\Gamma^1\\
&+\frac{A_2}{a_x}\sin(k_xa_x)\Gamma^3
+\frac{A_2}{a_y}\sin(k_ya_y)\Gamma^4+M_1(\mathbf{k_{\parallel}},k_z)\Gamma^5,\nonumber
\end{align}
with
\begin{align}
\epsilon_1(\mathbf{k_{\parallel}},k_z)\equiv &C+\frac{2D_1}{a_z^2}[1-\cos(k_za_z)]+\frac{2D_2}{a_x^2}[1-\cos(k_xa_x)]\nonumber\\
&+\frac{2D_2}{a_y^2}[1-\cos(k_ya_y)],
\label{eq:epsilon1}
\end{align}
\begin{align}
M_1(\mathbf{k_{\parallel}},k_z)\equiv &m-\frac{2B_1}{a_z^2}[1-\cos(k_za_z)]-\frac{2B_2}{a_x^2}[1-\cos(k_xa_x)]\nonumber\\
&-\frac{2B_2}{a_y^2}[1-\cos(k_ya_y)].
\label{eq:M1}
\end{align} 
The Gamma matrices are defined as $\Gamma^1=\sigma^z\otimes\tau^x$, $\Gamma^2=-\mathbb{I}_{2\times 2}\otimes\tau^y $, $\Gamma^3=\sigma^x\otimes\tau^x$, $\Gamma^4=\sigma^y\otimes\tau^x$, and $\Gamma^5=\mathbb{I}_{2\times 2}\otimes\tau^z$, where $\sigma^j$ and $\tau^j$ are Pauli matrices acting on $(\uparrow,\downarrow)$ and $(P_1,P_2)$ spaces, respectively.
$a_x$, $a_y$ and $a_z$ are the lattice constants in x, y, z directions respectively. 
We use the parameters for $\rm{Bi_2Se_3}$ obtained
by fitting the continuum model to 
ab-initio calculations\cite{Zhang2009}: $m=0.28$~eV, $A_1=2.2$~eV\r{A}, $A_2=4.1$~eV\r{A}, $B_1=10$~eV\r{A}$^2$,
$B_2=56.6$~eV\r{A}$^2$, $C=-0.0068$~eV, $D_1=1.3$~eV\r{A}$^2$ and $D_2=19.6$~eV\r{A}$^2$.

We model a slab of Bi$_2$Se$_3$ by imposing open boundary conditions at the top and the bottom surfaces which break the translational symmetry along the $z$-axis. 
Since $k_z$ is no longer a good quantum number, while $\mathbf{k_{\parallel}}$ still is, we substitute $c_{\mathbf{k_{\parallel}},k_z}=\frac{1}{\sqrt{N}}\sum_{j}e^{ik_zja_z}c_{\mathbf{k_{\parallel}},j}$ in Eq. (\ref{eq:H0}) and label the four-spinor operators by the in-plane momentum $\mathbf{k_{\parallel}}=(k_x,k_y)$ and the index of layers $j$ stacking in the z direction. Now the Hamiltonian for a slab with $N$ layers is
\begin{equation}
H_0(N)=\sum_{\mathbf{k_{\parallel}}}H_{0}(\mathbf{k_{\parallel}},N),
\label{eq:H0tb1}
\end{equation}
where
\begin{equation}
H_{0}(\mathbf{k_{\parallel}},N)=\sum_{j=1}^N\mathbb{M}~c^{\dagger}_{\mathbf{k_{\parallel}},j}c_{\mathbf{k_{\parallel}},j}
+\mathbb{T}c^{\dagger}_{\mathbf{k_{\parallel}},j+1}c_{\mathbf{k_{\parallel}},j}
+\mathbb{T}^{\dagger}c^{\dagger}_{\mathbf{k_{\parallel}},j}c_{\mathbf{k_{\parallel}},j+1}
\label{eq:H0tb2}
\end{equation}
and the $4\times 4$ matrices $\mathbb{T}$ and $\mathbb{M}$ are defined as  
\begin{equation}
\mathbb{T}\equiv -\frac{D_1}{a_z^2}\mathbb{I}_{4\times 4}+\frac{B_1}{a_z^2}\Gamma^5-\frac{iA_1}{2a_z}\Gamma^1,
\label{eq:T}
\end{equation}
and
\begin{equation}
\mathbb{M}\equiv \epsilon_2(\mathbf{k_{\parallel}})\mathbb{I}_{4\times 4}+\frac{A_2}{a_x}\sin(k_xa_x)\Gamma^3+\frac{A_2}{a_y}\sin(k_ya_y)\Gamma^4+M_2(\mathbf{k_{\parallel}})\Gamma^5
\label{eq:M}
\end{equation}
where
\begin{equation}
\epsilon_2(\mathbf{k_{\parallel}})\equiv C+\frac{2D_1}{a_z^2}+\frac{2D_2}{a_x^2}[1-\cos(k_xa_x)]+\frac{2D_2}{a_y^2}[1-\cos(k_ya_y)]
\label{eq:epsilon2}
\end{equation}
and
\begin{equation}
M_2(\mathbf{k_{\parallel}})\equiv m-\frac{2B_1}{a_z^2}-\frac{2B_2}{a_x^2}[1-\cos(k_xa_x)]-\frac{2B_2}{a_y^2}[1-\cos(k_ya_y)].
\label{eq:M2}
\end{equation}
{For the sake of simplicity, the results presented in the remainder of this section are calculated with $a_x=a_y=a_z=1\r{A}$.}\\

We can diagonalize $H_{0}(\mathbf{k_{\parallel}},N)$ as
\begin{equation}
\begin{split}
H_{0}(\mathbf{k_{\parallel}},N)=\sum_{\alpha=1}^{4N-4}E^0_{B,\alpha}(\mathbf{k_{\parallel}})b_{\alpha,\mathbf{k_{\parallel}}}^{0\dagger}b^0_{\alpha,\mathbf{k_{\parallel}}}\\
+\sum_{\beta=1}^{4}E^0_{D,\beta}(\mathbf{k_{\parallel}})d_{\beta,\mathbf{k_{\parallel}}}^{0\dagger}d^0_{\beta,\mathbf{k_{\parallel}}},
\end{split}
\label{eq:H0tbdiag}
\end{equation}
where $b^0_{\alpha,\mathbf{k_{\parallel}}}$ and $d^0_{\beta,\mathbf{k_{\parallel}}}$ are four-spinor annihilation operators for bulk and surface states (henceforth referred to as the ``Dirac" states) respectively in the absence of hybridization, $E^0_{B,\alpha}$ and $E^0_{D,\beta}$ are their corresponding eigenenergies.  Here, $\alpha$ and $\beta$ label the unhybridized bulk and Dirac states respectively. All energy eigenstates are 
two-fold degenerate as required by inversion (P) and time-reversal (T) symmetries. For each  in-plane momentum $\mathbf{k_{\parallel}}$, four energy eigenstates with their energies closest to the Dirac point are labeled to be valence ($\beta=1,2$) and conduction ($\beta=3,4$) Dirac states. $\alpha$ label the remaining $4N-4$ bulk states. A natural choice for the labeling is to let $\alpha=1, \cdots, 2N-2$ denote valence bulk states  and let $\alpha=2N-1,\cdots, 4N-4$ denote conduction bulk states
with the eigenenergies increasing monotonically with $\alpha$.
As the model is derived from a low-energy effective model near the Dirac point, it will break down at large energies. However, we expect qualitatively correct results when it comes to trends of physical properties over the hybridization strength and film thickness which only require knowledge of the low-energy physics near the insulating gap.  

Now we introduce the S-B hybridization term that is allowed by symmetries in the spirit of Fano model\cite{Mahan}.
The Fano model is a generic model describing the mixing between extended states $c_k$ with energy $\epsilon_k$ and a localized state $b$ with energy $\epsilon$ through Hamiltonian 
 $H_F=\epsilon b^{\dagger}b+\sum_k[\epsilon_kc^{\dagger}_kc_k+A_k(c^{\dagger}_kb+b^{\dagger}c_k)]$, where  $A_k$ represents scattering strength. 
\textcite{PhysRevB.82.195417} pointed out that $H_F$ can be used to describe the lowest order interaction between a helical surface state and a metallic bulk band, i.e. hybridization. They studied effects of hybridization in a field theoretic approach. Here we use a symmetry-preserving form of the surface-bulk hybridization term in the spirit of $H_F$ for the microscopic model of 
 $\rm{Bi_2Se_3}$ shown in Eq.~\eqref{eq:H0tbdiag} to study the effects of mixing between Dirac states and bulk states. 

For simplicity we consider the case where the hybridization strength preserves in-plane momenta $\mathbf{k_{\parallel}}$ and is independent of energy and $\mathbf{k_{\parallel}}$, i.e. 
\begin{equation}
h'(\mathbf{k_{\parallel}})=\sum_{\alpha,\beta}g\; b_{\alpha,\mathbf{k_{\parallel}}}^{0\dagger}d^0_{\beta,\mathbf{k_{\parallel}}}+H.c..
\label{eq:hhyb}
\end{equation}
We then impose $T$ and $P$ symmetries on the 
full hybridization perturbation 
$H'(\mathbf{k_{\parallel}})$
by constructing $H'(\mathbf{k_{\parallel}})$ through
\begin{align}
H'(\mathbf{k_{\parallel}})&=h'(\mathbf{k_{\parallel}})+Ph'(\mathbf{k_{\parallel}})P^{\dagger}+Th'(\mathbf{k_{\parallel}})T^{\dagger}\nonumber\\
&+PTh'(\mathbf{k_{\parallel}})(PT)^{\dagger},
\label{eq:Hhyb}
\end{align}
where the representations for $T$ and $P$ symmetry operators with the spatial inversion center at the middle point of the slab in the current $4N$ tight-binding spin-orbital basis are $T=K\; i\sigma^y\otimes \mathbb{I}_{2\times 2} \otimes$ $\mathbb{I}_{N\times N}$ with $\mathbf{k_{\parallel}}\leftrightarrow -\mathbf{k_{\parallel}}$, and $P=\mathbb{I}_{2\times 2}\otimes\tau^z$ with $z: [0,N/2]\leftrightarrow[N/2,N]$ and $\mathbf{k_{\parallel}}\leftrightarrow -\mathbf{k_{\parallel}}$, respectively. Here, $K$ is the usual complex conjugation operator. Finally, the full Hamiltonian including hybridization at a given in-plane momentum $\mathbf{k_{\parallel}}$ reads
\begin{align}
H(\mathbf{k_{\parallel}})=H_0(\mathbf{k_{\parallel}},N)+H'(\mathbf{k_{\parallel}}).
\label{eq:Htot}
\end{align} 
After diagonalizing the full Hamiltonian in the tight-binding spin-orbital bases, we can write
\begin{equation}
\begin{split}
H(\mathbf{k_{\parallel}})=\sum_{\alpha=1}^{4N-4}E_{B,\alpha}(\mathbf{k_{\parallel}})b_{\alpha,\mathbf{k_{\parallel}}}^{\dagger}b_{\alpha,\mathbf{k_{\parallel}}}\\
+\sum_{\beta=1}^{4}E_{D,\beta}(\mathbf{k_{\parallel}})d_{\beta,\mathbf{k_{\parallel}}}^{\dagger}d_{\beta,\mathbf{k_{\parallel}}},
\end{split}
\label{eq:Htotdiag}
\end{equation}
where $b_{\alpha,\mathbf{k_{\parallel}}}$, $d_{\beta,\mathbf{k_{\parallel}}}$, $E_{B,\alpha}$ and $E_{D,\beta}$ are defined similarly to the corresponding symbols with a superscript $0$ in Eq.~\eqref{eq:H0tbdiag} but in the presence of hybridization. $E_{B,\alpha}$ and $E_{D,\beta}$ are again two-fold degenerate as $H(\mathbf{k_{\parallel}})$ preserves parity and time-reversal symmetries by design. $\alpha$ and $\beta$ label the bulk and Dirac states for $H(\mathbf{k_{\parallel}})$, where the terms bulk and Dirac states are defined in the same fashion as for $H_{0}(\mathbf{k_{\parallel}},N)$ in the absence of hybridization.

\subsection{Topological Metal Regime}
\begin{figure}[h]
\centering
\subfigure[]{
	\includegraphics[width=4cm]{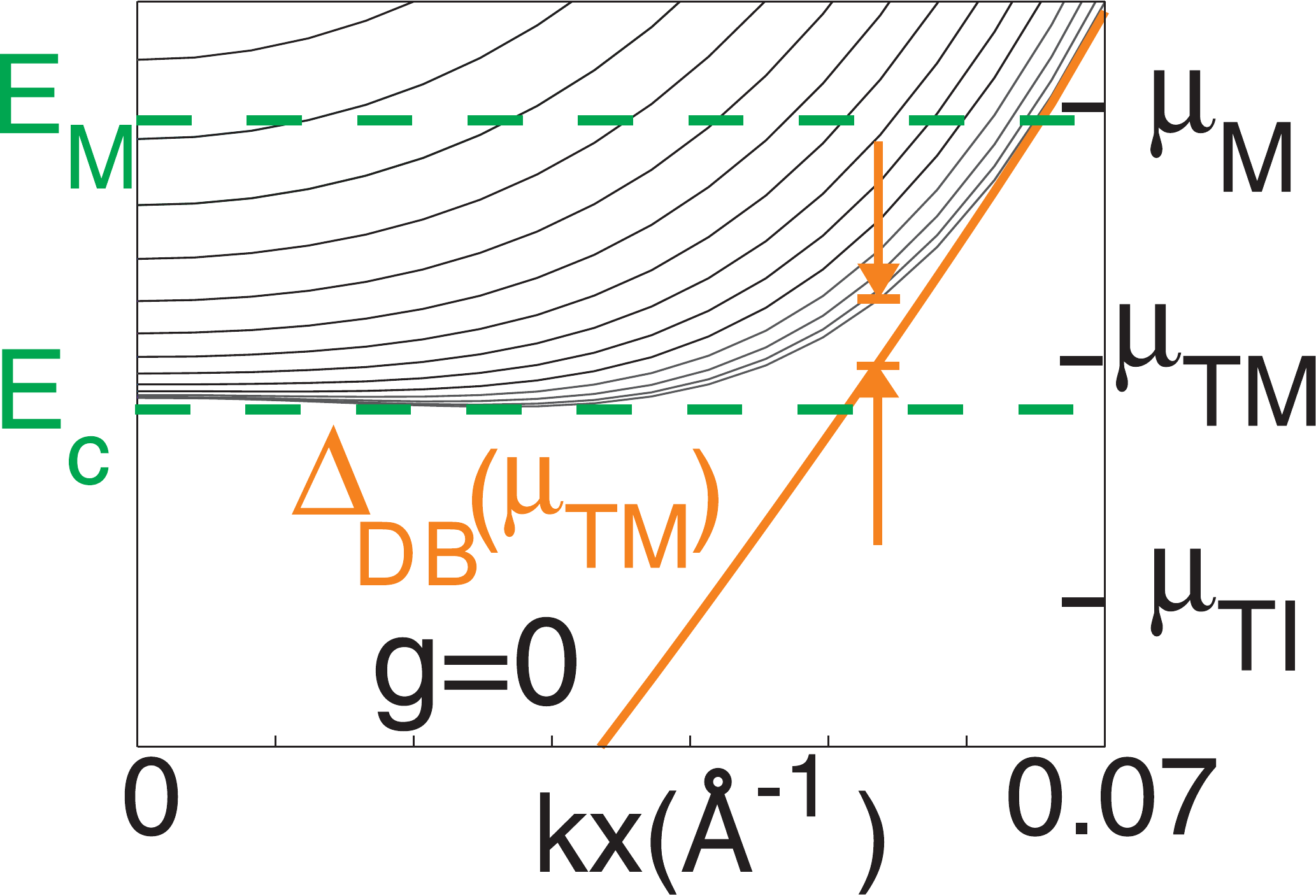}
}
\subfigure[]{
	\includegraphics[width=4cm]{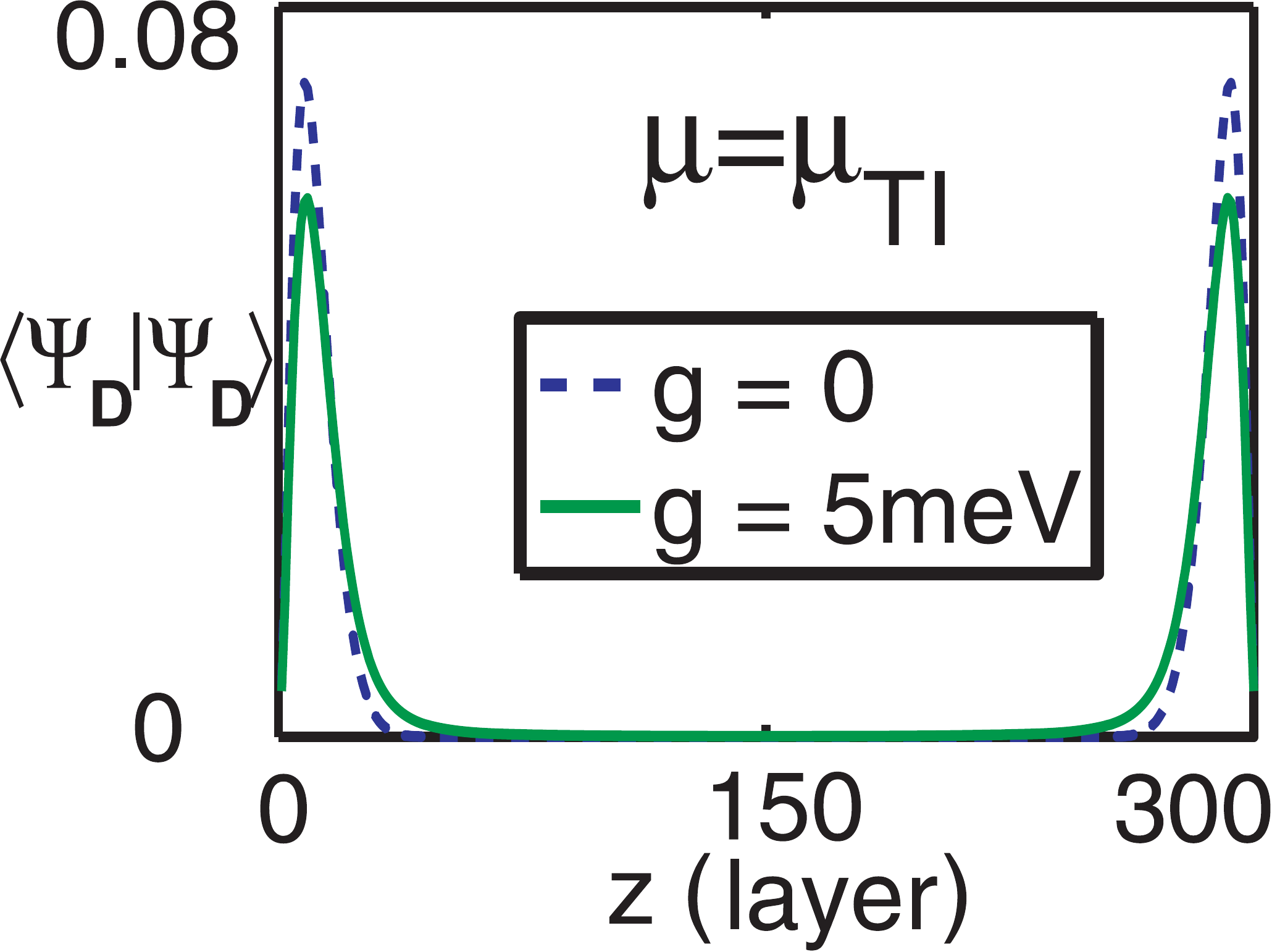}
}
\subfigure[]{
	\includegraphics[width=4cm]{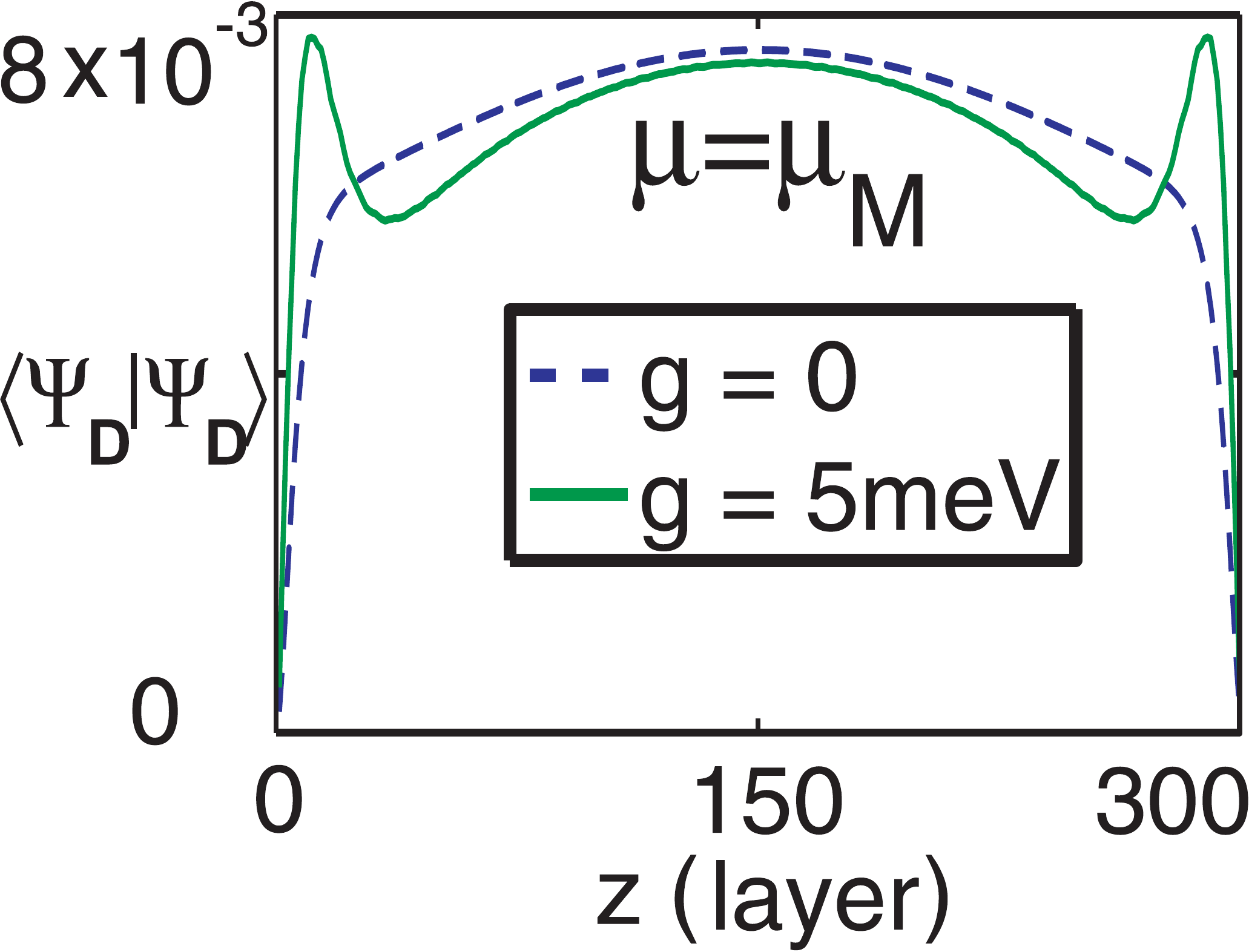}
}
\subfigure[]{
	\includegraphics[width=4cm]{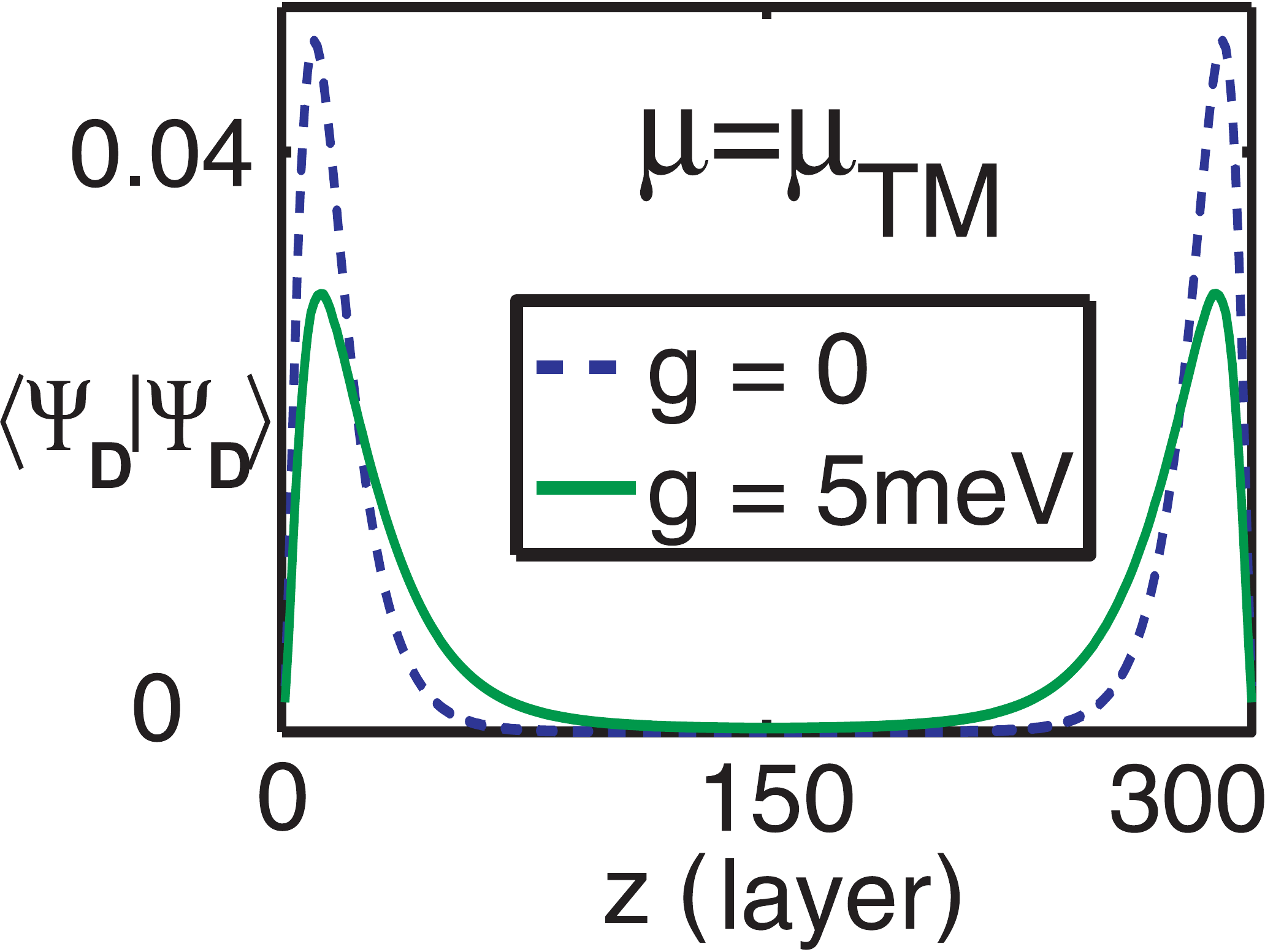}
}
\subfigure[]{
	\includegraphics[width=4cm]{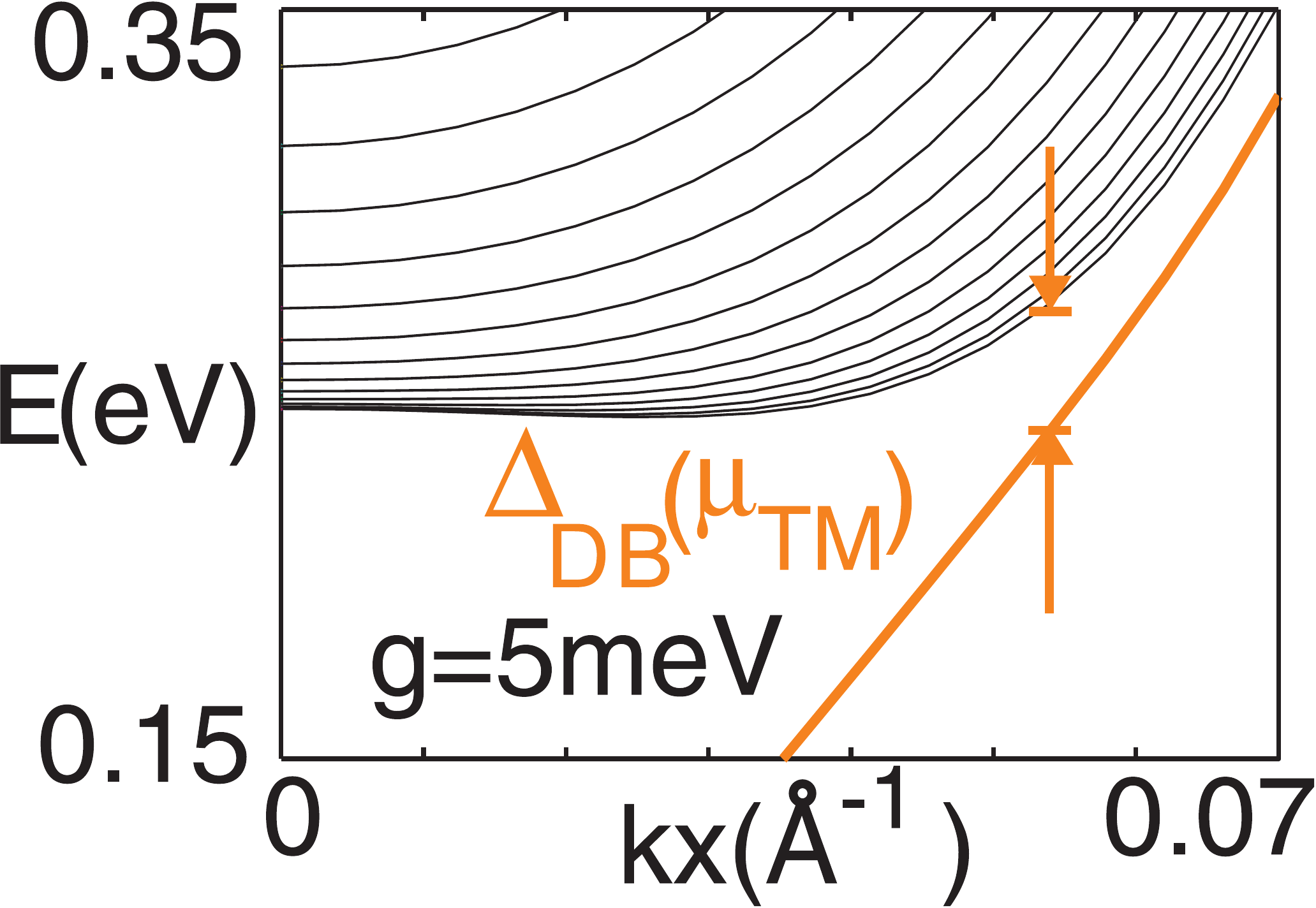}
}
\subfigure[]{
	\includegraphics[width=4cm]{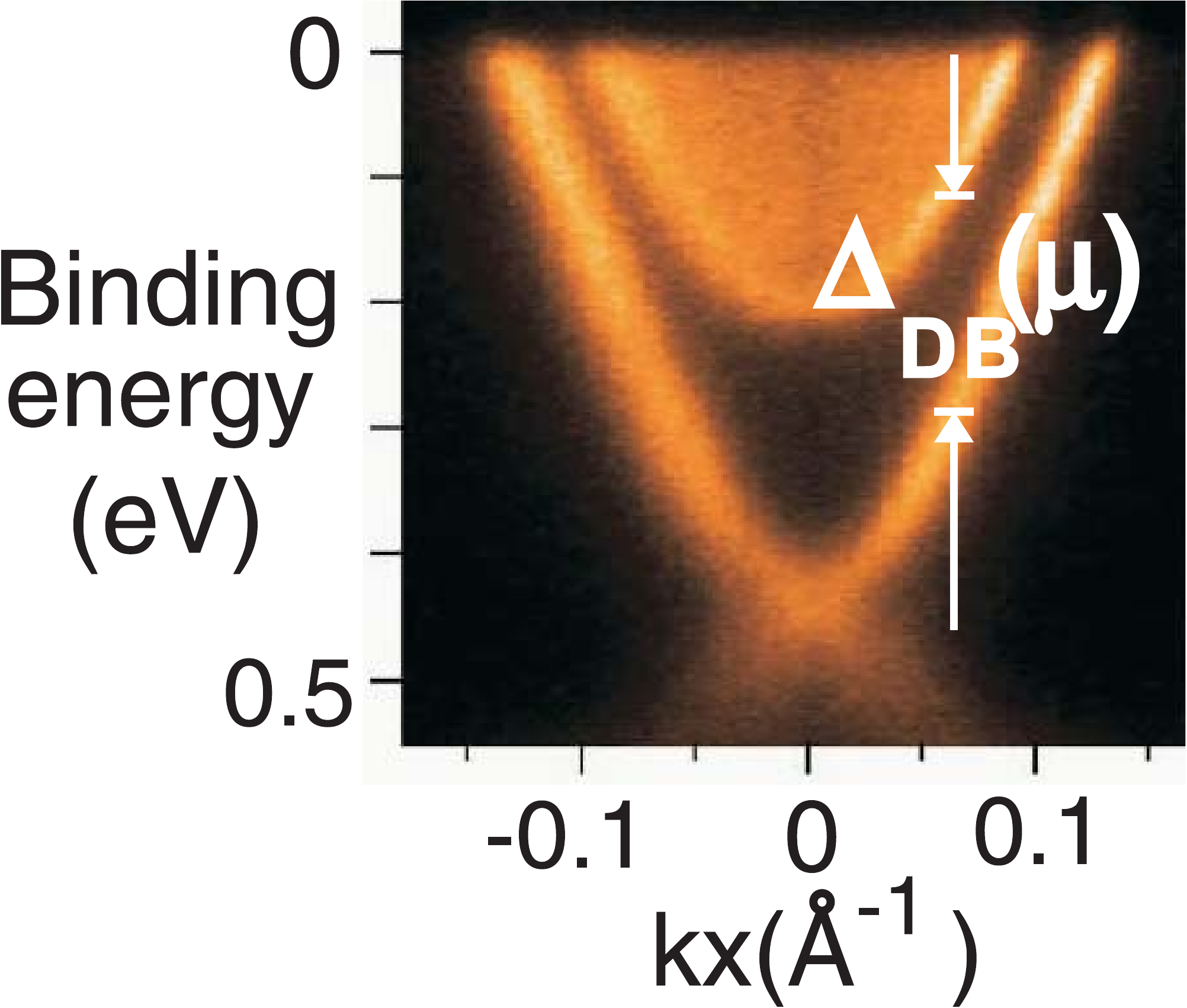}
}
\caption{(a) Spectra of the model on a 300-layer-thick slab.
The three chemical potentials $\mu_{TI}$, $\mu_{TM}$ and $\mu_{M}$ are taken as 
representative points for the three regimes TI($\mu\leq {E_c}$), 
M ($\mu\gtrsim{E_M}$), and TM(${E_c}\leq\mu\lesssim{E_M}$) as defined in the text, respectively. (b)-(d)The corresponding unhybridized(dashed, blue) and hybridized(solid, green) spatial profiles of the pair of degenerate conduction Dirac states  $|\Psi_{D,\mathbf{k}_\parallel}(z)|^2$ at $\mathbf{k}_\parallel=\mathbf{k}_{\parallel,\mu}$ with $\mu=\mu_{TI}$, $\mu_{TM}$, and $\mu_{M}$, respectively. (e) Effect of the hybridization on the spectra. 
 (f)ARPES data on Bi$_2$Se$_3$\cite{Bianchi2010}.}
\label{spectrum}
\end{figure}

We begin our numerical study with no hybridization. In the absence of hybridization, depending on the chemical potential $\mu$, we now define three regimes: topological insulator(TI), metal(M), and topological metal(TM)(see Fig.~\ref{spectrum}(a)). The familiar topological insulator (TI) regime is where the chemical potential lies within the bulk gap and the system is actually a bulk band insulator, i.e., 
$E_v \leq \mu\leq {E_c}$ with ${E_c}$ being the bottom of the conduction band and $E_v$ the top of the valence band.
Within the TI regime, the Dirac states feature Rashba-type spin-momentum locking and a spatial profile localized on the surfaces.
Of our particular interest is the distinction we will draw between M and TM regimes based on whether the Dirac states retain the spin-momentum locking and the surface localization when being away from the TI regime. 

In order to examine the above two properties of Dirac states, we 
define
$|\psi_{D,\mathbf{k_{\parallel}}}\rangle\equiv d_{3,\mathbf{k_{\parallel}}}^{\dagger}|0\rangle\ (d_{3,\mathbf{k_{\parallel}}}^{0\dagger}|0\rangle)$ and $|\tilde{\psi}_{D,\mathbf{k_{\parallel}}}\rangle\equiv d_{4,\mathbf{k_{\parallel}}}^{\dagger}|0\rangle\ (d_{4,\mathbf{k_{\parallel}}}^{0\dagger}|0\rangle)$ in the presence(absence) of hybridization to represent the pair of degenerate Dirac states above the Dirac point. 
Now the spatial profile of the conduction Dirac states is $|\Psi_{D,\mathbf{k}_{\parallel}}(z)|^2\equiv|\psi_{D,\mathbf{k}_{\parallel}}(z)|^2+|\tilde{\psi}_{D,\mathbf{k}_{\parallel}}(z)|^2$ which is a function of $z$ measured from the bottom of the slab along the finite dimension of the slab. This quantity will show whether the Dirac states are localized on the surfaces or not. Let us identify a particular  $\mathbf{k}_{\parallel}$ of interest for a given value of chemical potential as the in-plane momentum at which the chemical potential $\mu$ intersects the Dirac branch; we denote such in-plane momentum by 
$\mathbf{k}_{\parallel,\mu}$. To illustrate the features defining the three regimes, we will now show the spatial profiles and the spin polarizations of the Dirac states in the three regimes in the absence of hybridization. In the next section we will add the hybridization and examine its effects.

Inspecting $|\Psi_{D,\mathbf{k}_{\parallel}}(z)|^2$ at $\mathbf{k}_{\parallel}=\mathbf{k}_{\parallel,\mu}$ at the representative values of chemical potential for the three regimes $\mu_{TI}, \mu_{M}, \mu_{TM}$ shown in Fig.~\ref{spectrum}, we find that 
the spatial profile of the Dirac states $|\Psi_{D,\mathbf{k}_{\parallel}}(z)|^2$ indicates surface localized states of the slab in the TI regime as expected (see Fig.~\ref{spectrum}(b)).  On the other hand, in the M regime, where the chemical potential is well within the bulk conduction band, $|\Psi_{D,\mathbf{k}_{\parallel}}(z)|^2$ is 
fully delocalized over the entire slab (see Fig.~\ref{spectrum}(c)). In this regime, the system cannot be distinguished from an ordinary metal.
However, even with $\mu>E_c$ there is an 
energy window between
$E_c$ and a crossover energy scale $E_M$, where the Dirac states are still spatially localized on the surfaces in the sense that $|\Psi_{D,\mathbf{k}_{\parallel}}(z)|^2$ is peaked on each surface of the slab and decays away from the surfaces (see Fig.~\ref{spectrum}(d)).
The crossover energy scale $E_M$ is a threshold energy, where, within the regime $\mu\gtrsim E_M$ (regime M), wavefunctions for all states at in-plane momentum $\mathbf{k}_{\parallel,\mu}$ delocalize. 
We define the system to behave as a topological metal(TM) when the chemical potential lies within this window, i.e. $E_c<\mu<E_M$, represented by $\mu_{TM}$. 

\begin{figure}[h]
\centering
\subfigure[]{
	\includegraphics[width=4cm]{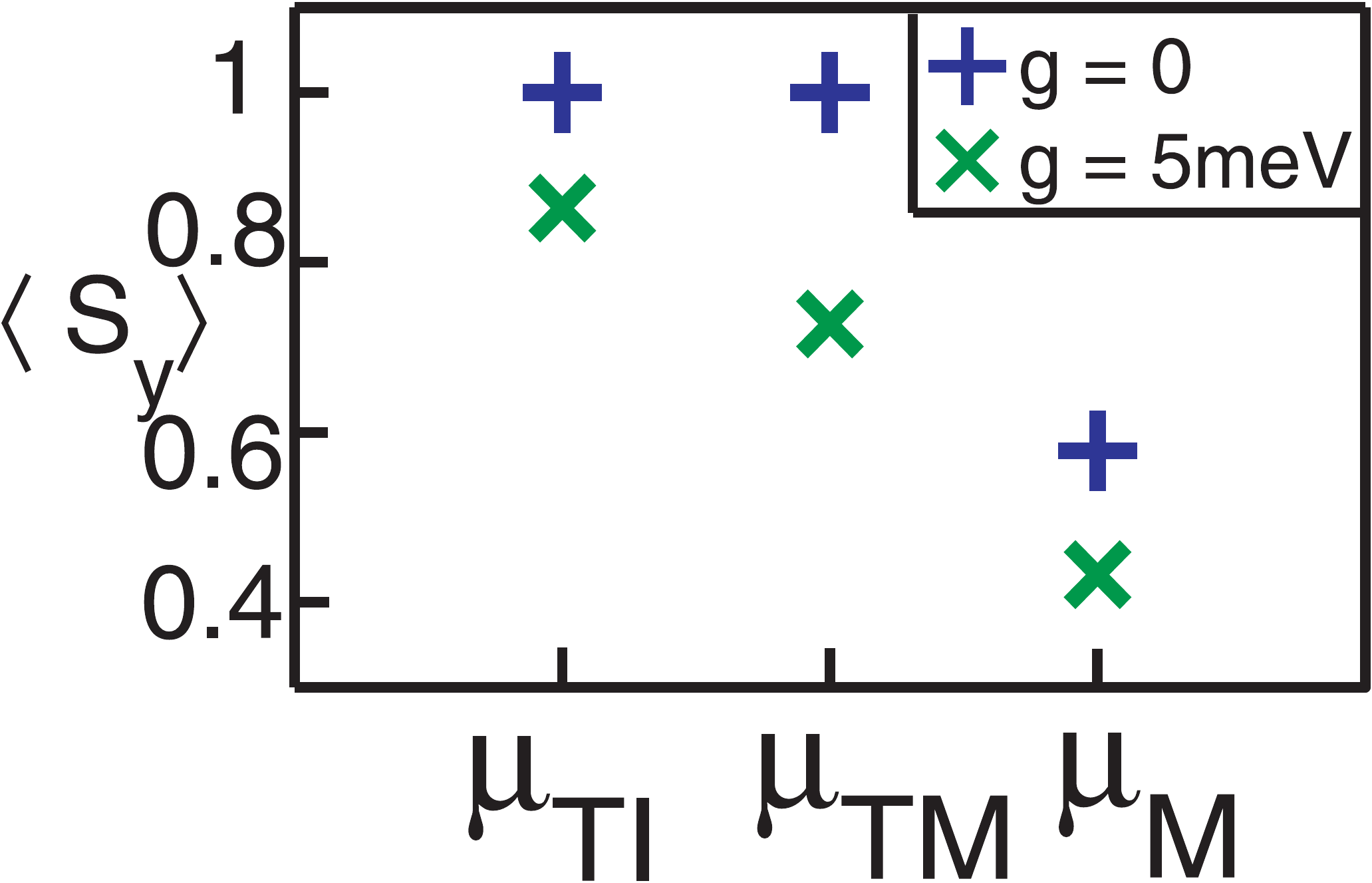}
	}
\subfigure[]{
	\includegraphics[width=4cm]{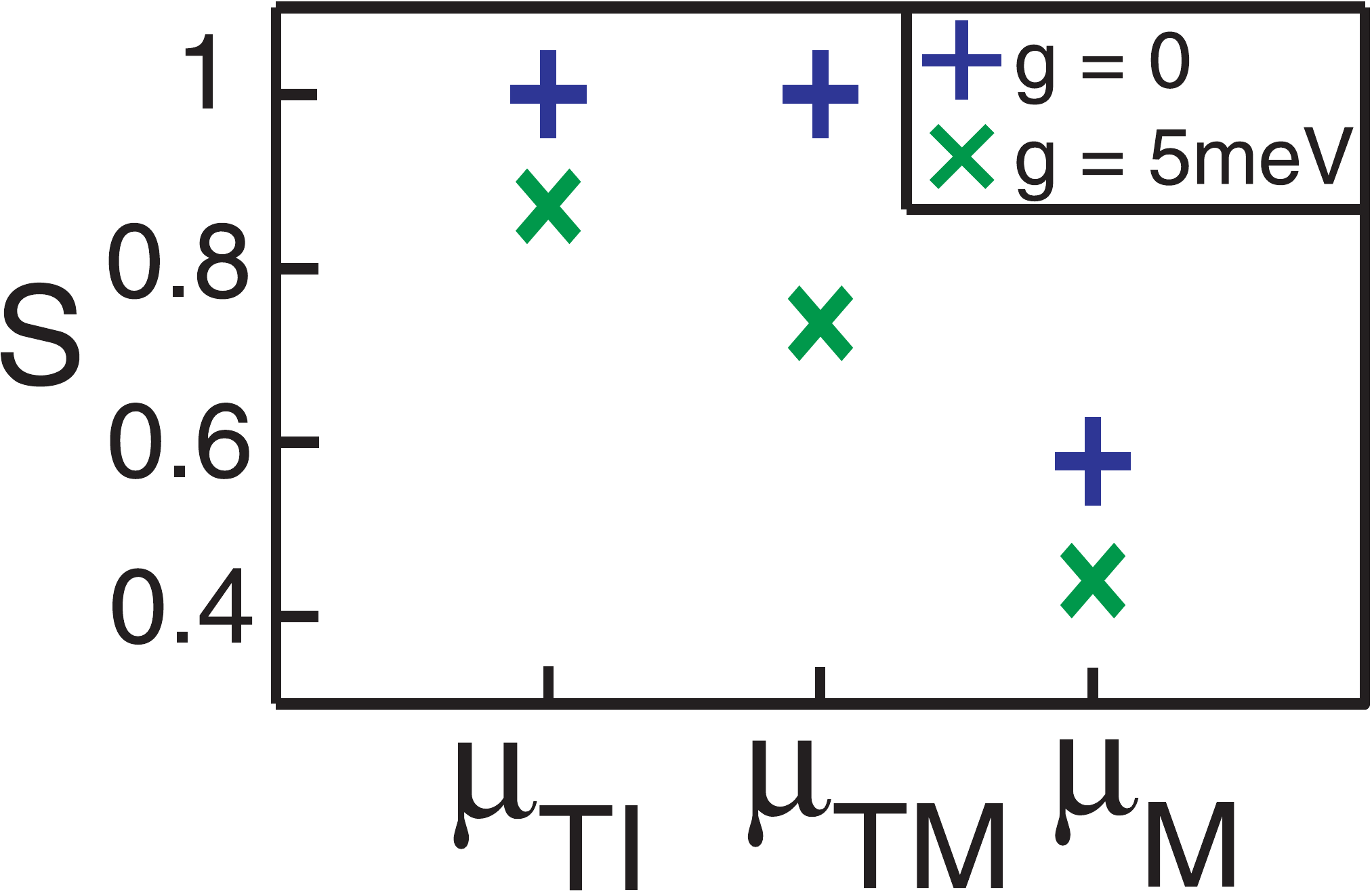}
	}
\caption{The effect of hybridization on the degree of spin-momentum locking in different regimes. The spin expectation values are calculated for a 300-layer-thick slab using a conduction Dirac state $|\psi_{D,k_x\hat{x}}\rangle$ with $k_x=k_{x,\mu}$ at different representative chemical potentials $\mu=\mu_{TI}$, $\mu_{TM}$ and $\mu_{M}$. 
(a)Spin polarization $\langle S_y\rangle (k_x\hat{x})$. (b)Total spin magnitude $S(k_x\hat{x})$(defined in the text).
}
\label{spin}
\end{figure}

A detectable characteristic of TI and TM regimes is the spin-momentum locking. 
One measure to quantify spin-momentum locking at the surface is through the so-called ``spin polarization'' which is the  
expectation value of the spin component 
perpendicular to the in-plane momentum of a Dirac state, i.e., 
\begin{equation}
 \langle S_{\hat{\mathbf{n}}}\rangle(\mathbf{k_{\parallel}})\equiv\langle\psi_{D,\mathbf{k}_\parallel}|S_{\hat{\mathbf{n}}}|\psi_{D,\mathbf{k}_\parallel}\rangle
\label{eq:Sn}
\end{equation}
 with $\mathbf{\hat{n}}\cdot \mathbf{k_{\parallel}}=0$. 
Here the $\hat{\mathbf{n}}$ component of the quantum spin operator is defined as $S_{\hat{\mathbf{n}}}\equiv \frac{\hbar}{2}\vec{\sigma}\cdot\mathbf{\hat{n}}\otimes \mathbb{I}_{\rm{2x2}}\otimes \mathbb{I}_{\rm{NxN}}$, 
where $\vec{\sigma}=(\sigma_x,\sigma_y,\sigma_z)$ are Pauli matrices acting on spin, $\mathbb{I}_{\rm{2x2}}$ acts on the  orbital degree of freedom , and $\mathbb{I}_{\rm{NxN}}$ acts on the layer index.
$\langle S_y\rangle(k_x\hat{x})$ is evaluated at $k_x\hat{x}=k_{x,\mu}\hat{x}$ and shown for $\mu=\mu_{TI}, \mu_{TM}, \mu_M$ in 
Fig.~\ref{spin}(a). We see here that, in the absence of hybridization(g=0), the spin polarization stays maximal in the TI and TM regimes while rapidly dropping upon entering the M regime. 
Another quantity of experimental interest is %
the total spin 
magnitude associated with the Dirac states with in-plane momentum $\mathbf{k}_\parallel$ defined in terms of spin polarization as %
\begin{equation}
S(\mathbf{k_{\parallel}})\equiv\sqrt{\sum_{i=x,y,z}\left[\langle S_i\rangle (\mathbf{k_{\parallel}})\right]^2}. 
\label{eq:spin-mag}
\end{equation}
Fig.~\ref{spin}(a) and \ref{spin}(b) show that the spin-momentum locking quantified using these measures clearly distinguishes the TM regime from the ordinary metal regime (M) in the absence of hybridization.

\subsection{Effects of S-B Hybridization}
We now turn to the effects of hybridization. One effect of hybridization that is manifest in the experimental detection of Dirac surface states in the TM regime is an increase in the bulk-Dirac state energy gap. We quantify this energy gap, for a given chemical potential $\mu,$ using the energy difference between a Dirac state above the Dirac point and the energetically closest bulk state defined by 
\begin{equation}
\Delta_{DB}(\mu)\equiv E^{(0)}_{B,2N-1}(k_{\parallel,\mu})-E^{(0)}_{D,3}(k_{\parallel,\mu})
\end{equation}
in the presence(absence) of hybridization. 
Comparing Fig.~\ref{spectrum}(a) to (e), we find that the key effect of hybridization that is spectroscopically detectable is 
the increase in $\Delta_{DB}(\mu)$ in both TM and M regimes compared to the TI regime.  
Otherwise the spectra in the absence or presence of hybridization look similar. 
Note that most ARPES data on 3D TIs exhibit a clear energy gap between the Dirac branch and the bulk states at a chemical potential well into the bulk band as shown in Fig.~\ref{spectrum}(f). This experimental trend hints at the possibility that a sizable hybridization between Dirac states and the bulk states is common in 3D TI materials. In order to demonstrate the effect of hybridization, we choose a value of $g=5$meV that is subdominant to all the hopping terms yet substantial in this paper. However, key effects of hybridization do not depend qualitatively on the value of $g$.

Another effect of hybridization is to broaden the Dirac state wavefunctions in the TI and TM regimes.  The degree of broadening depends on 
the chemical potential $\mu$, hybridization strength $g$, and the slab thickness $N$. 
However, as long as $g$ is the smallest energy scale in the total Hamiltonian as is the case for Figs.~\ref{spectrum}(b-d), the Dirac states in the TI and TM regimes remain localized on the surfaces. A tangible consequence of the wavefunction broadening is the quantitative suppression of the spin-momentum locking. 
As mentioned earlier, in the absence of hybridization the Dirac states of TI and TM exhibit a maximal degree of spin-momentum locking.   However, hybridization rotates the spin vectors of different atomic orbitals and layers away from the direction perpendicular to the in-plane momentum.  Hence, both measures of spin-momentum locking shown in Fig.~\ref{spin} show quantitative reduction upon hybridization. This is in qualitative agreement with the low values of spin polarization and total spin magnitude found in a first-principle calculation of a thin slab in a previous work\cite{PhysRevLett.105.266806} and our DFT results in the next section.  Note that the hybridization still preserves the spin-texture winding despite the quantitative reduction in the spin polarization.

\begin{figure}[t]
\centering
\subfigure[]{
	\includegraphics[width=4cm]{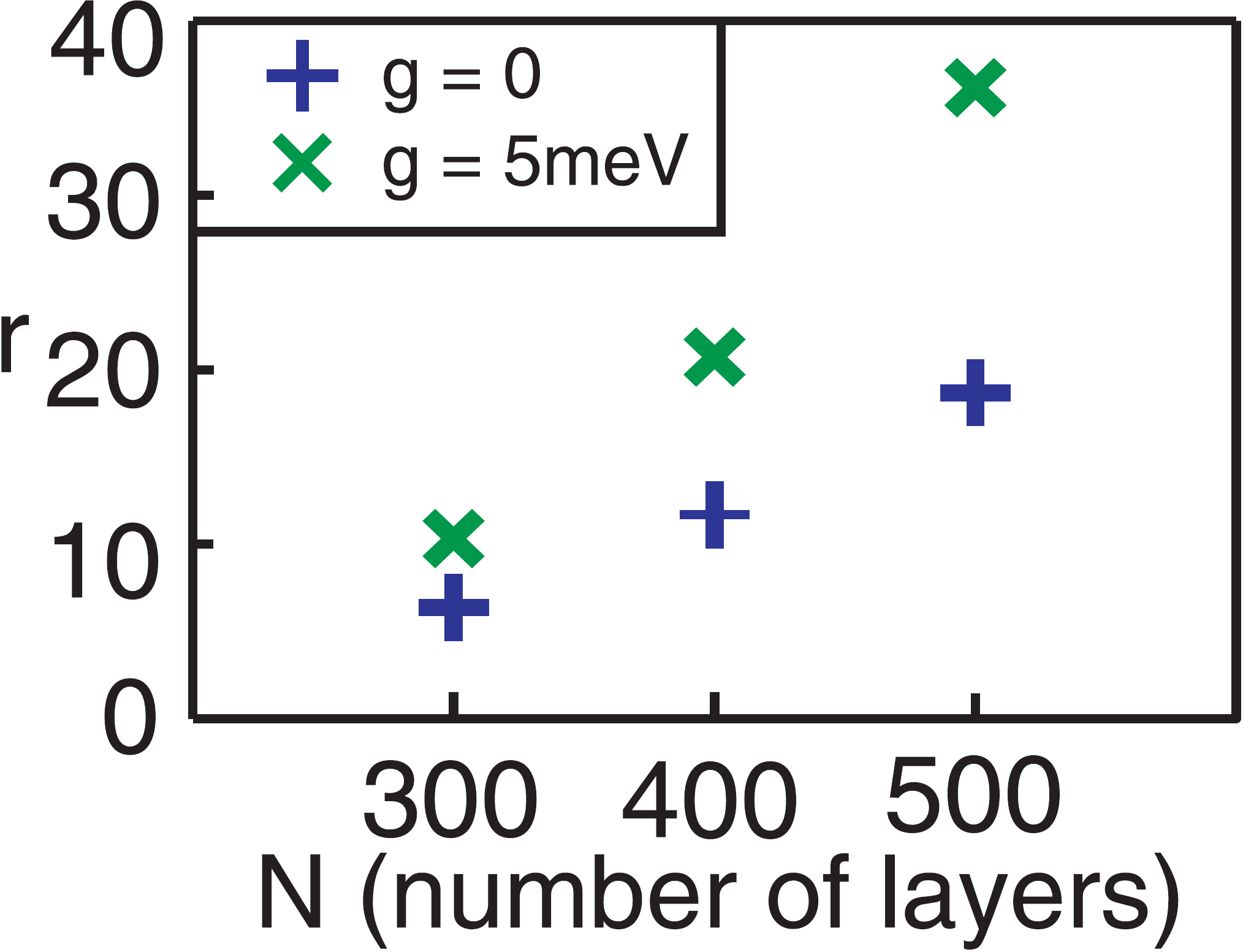}
	\label{fig:subfig3a}}
\subfigure[]{
	\includegraphics[width=4cm]{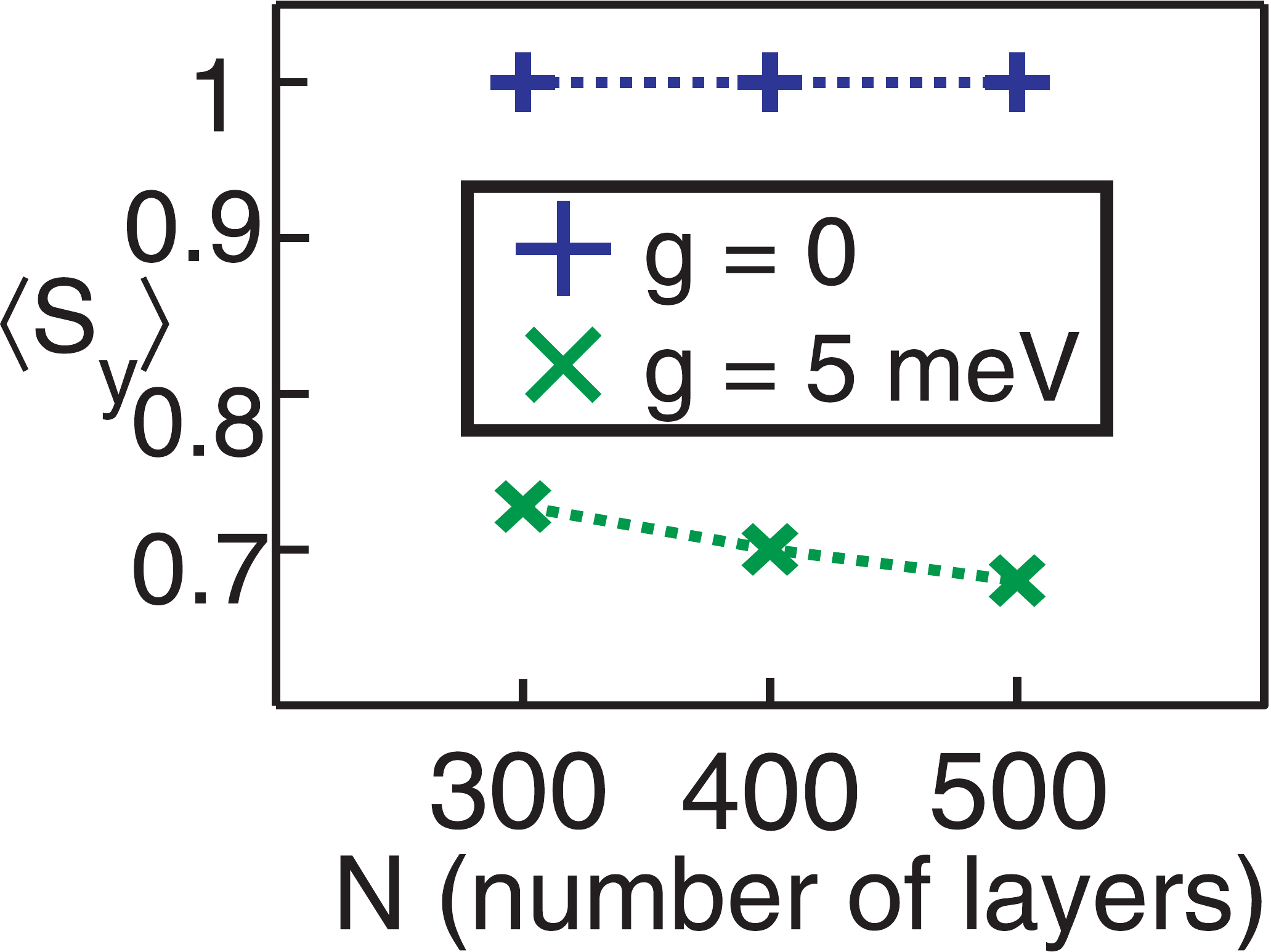}
	\label{fig:subfig3b}}
\caption{Thickness dependence of the hybridization effects.
(a)Dimensionless measure of bulk-Dirac energy gap $r=\Delta_{DB}/\Delta_{BB}$(defined in the text) at different slab thickness for $\mu=\mu_{TM}$. 
(b)The spin polarization of a conduction Dirac state $\langle S_y\rangle(k_x\hat{x})$ at $\mu=\mu_{TM}$.}
\label{tbNdep}
\end{figure}

Finally, we study how the effects of hybridization on the two experimentally accessible characteristics of the TM regime, namely how the bulk-Dirac energy gap $\Delta_{DB}(\mu)$ and the spin polarization $\langle S_{\hat{n}}\rangle(\mathbf{k}_\parallel)$ of a Dirac state, vary with the slab thickness.
Since the quantized energy spacings due to finite size effects decreases with increasing slab thickness, we consider a dimensionless measure that quantifies the bulk-Dirac energy gap:
\begin{equation}
r(\mu)\equiv\Delta_{DB}(\mu)/\Delta_{BB}(\mu),
\label{eq:hhyb}
\end{equation}
where $\Delta_{BB}(\mu)\equiv E^{(0)}_{B,2N+1}(k_{\parallel,\mu})-E^{(0)}_{B,2N-1}(k_{\parallel,\mu})$ is the energy spacing in the presence(absence) of hybridization between the two lowest lying conduction bulk branches measured at the same in-plane momentum $\mathbf{k_{\parallel,\mu}}$ where $\Delta_{DB}(\mu)$ is calculated. This dimensionless quantity $r(\mu)$ allows us to compensate for finite size effects though $\Delta_{BB}$ would be hard to measure experimentally for realistic bulk samples due to the lack of the required energy resolution.
Fig.~\ref{tbNdep}(a) shows that the hybridization induced enhancement in the bulk-Dirac energy gap becomes more prominent with increasing slab thickness. 
Comparing the existing ARPES data on bulk samples\cite{Bianchi2010} and on thin films\cite{Zhang2010}, we find the Dirac branch to be better separated from the bulk states in the bulk samples than in the thin films, which is consistent with the hybridization effect shown in Fig.~\ref{tbNdep}(a). 
Finally Fig.~\ref{tbNdep}(b) shows that the reduction in spin-polarization magnitude $|\langle S_y\rangle(k_x\hat{x})|$ is also intensified with increasing thickness. 
Such an enhancement in the impact of hybridization with the increase in slab thickness can be explained from the fact that a thicker slab implies a larger number of bulk states that mix with a fixed number of Dirac surface states for a given strength of hybridization $g.$

\section{DFT calculations of thin Bi$_2$Se$_3$ slabs} 
 \label{sec:DFT}

Now we turn to an ab-initio study of thin slabs to compare with the simple phenomenological model of hybridization we explored in the previous section. The approach of the previous section is limited, in the sense that it builds on a low-energy effective description of the band structure, and that there is no detailed knowledge of the hybridization strength $g$ which could in principle be $\mathbf{k}_\parallel$-dependent.  On the other hand, the DFT approach on slabs, which does not require calculating surface and bulk separately as in the calculations of semi-infinite systems\cite{Zhang2009}, is limited to very thin films of several QLs due to computational limits. By combining the two approaches, we extract a more robust understanding of the effects of hybridization in the TM regime and implications on their trends over film thickness.

We calculate the electronic structure of Bi$_2$Se$_3$(111) slabs of 4-6 QLs using the {\tt VASP} code 
\cite{VASP1,VASP2} with the projector-augmented-wave method \cite{PAW}, within the generalized-gradient 
approximation (GGA) \cite{PERD96}. Spin-orbit coupling is included self-consistently. We use experimental lattice constants \cite{NAKA63} and an energy cutoff of 420 eV with a $31 \times 31 \times 1$ $k$-point grid. Our DFT 
calculations are limited up to 6 QLs. For 5-6 QLs, the overlap
between top and bottom surface states is already very small yielding an energy gap of the order of meV 
at $\Gamma$. Expectation values of spin components $\langle S_x\rangle$, $\langle S_y\rangle$, $\langle S_z\rangle$ are calculated from the summation of the
expectation values of each atom.

\begin{figure}[h]
\includegraphics[width=8cm]{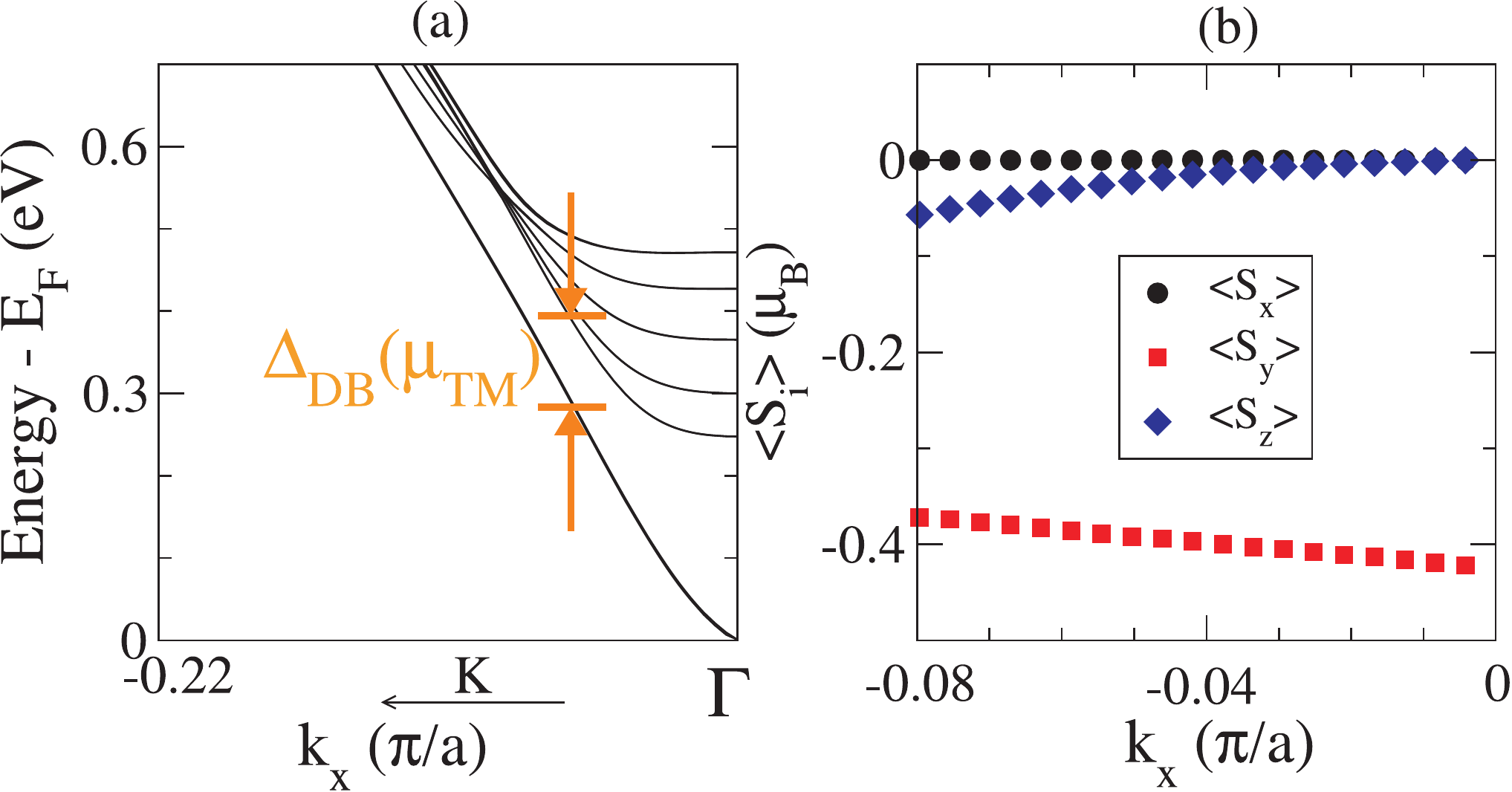}
\caption{(a) DFT-calculated band structure of a 6-QL slab of $\rm{Bi_2Se_3}$. (b)DFT-calculated
spin expectation values of the conduction Dirac state $\langle S_i\rangle(k_x\hat{x})$  for a 6-QL $\rm{Bi_2Se_3}$ slab.}
\label{DFT}
\end{figure}
\begin{figure}[h]	
\includegraphics[width=8cm]{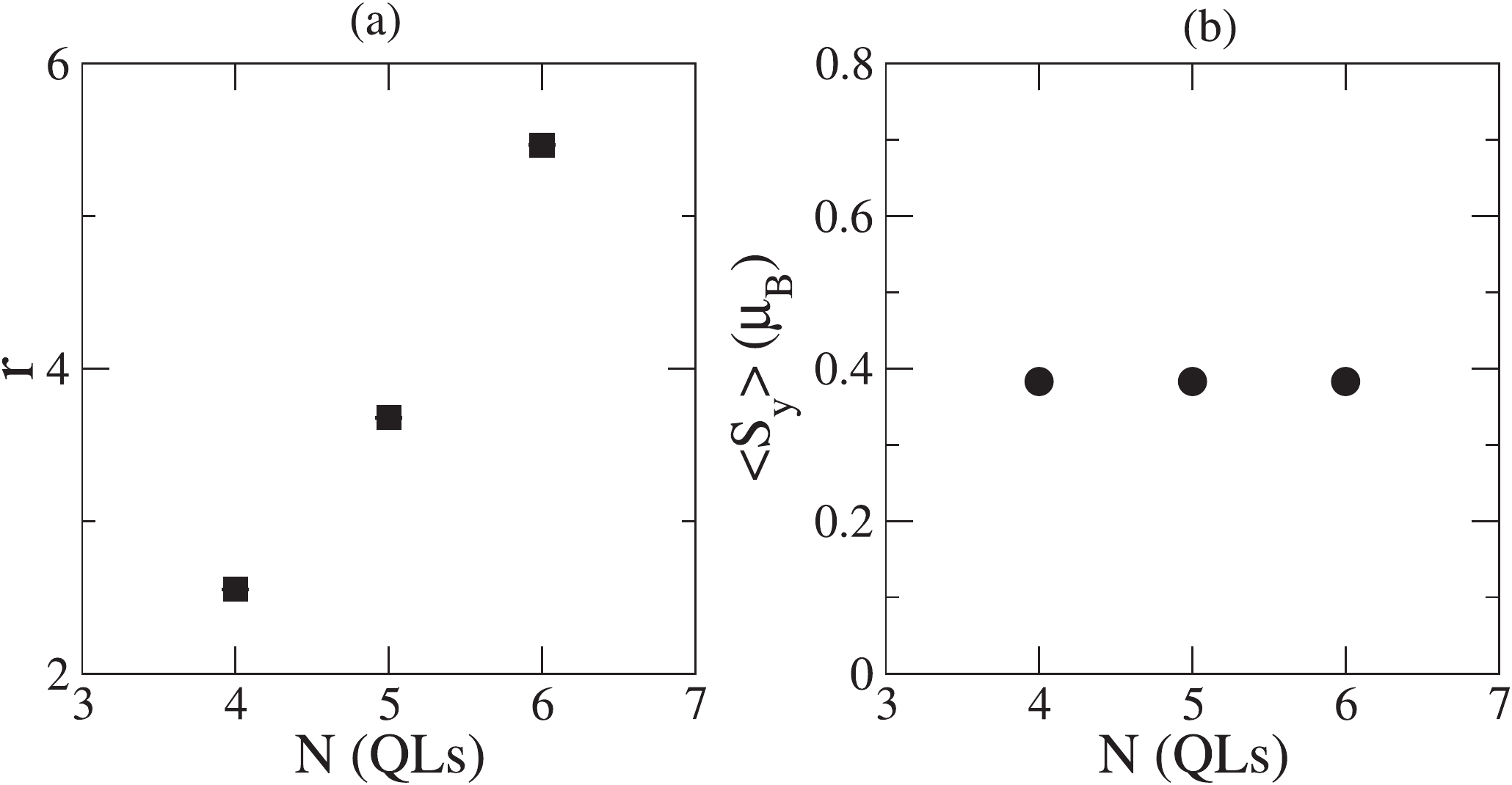}
\caption{(a) Ratio $r=\Delta_{DB}/\Delta_{BB}$ within the TM regime, calculated at a fixed $\mathbf{k}_\parallel$. 
 (b) $\langle S_y\rangle$ of the conduction Dirac state calculated using DFT as a function of 
slab thickness $N$.}
\label{DFTNdep}
\end{figure}

Figure~\ref{DFT} shows the DFT-calculated band structure and spin expectation values $\langle S_i\rangle(k_x\hat{x})$ of a 6-QL slab. 
The surface states are doubly degenerate and have a Dirac dispersion and we show five confined states
in the bulk conduction band region [Fig.~\ref{DFT}(a)]. 
For small $|k_x|$ values,
$\langle S_y\rangle$ of a Dirac conduction state is clearly dominant over other components and exhibits spin-momentum locking [Fig.~\ref{DFT}(b)]. As $|k_x|$ increases, a small 
$z$ component of spin expectation value develops.
 However, over the entire range of $k_x$, $\langle S_y\rangle$ is much less than the maximal value, in agreement with previous DFT study\cite{PhysRevLett.105.266806}. A comparison between Fig.~\ref{spin} and Fig.~\ref{DFT}(b) indicates that our hybridization model is an effective way to capture the broadening of the Dirac surface state wavefunction and the resulting reduction in the spin polarization and the total spin magnitude\footnote{Our  DFT calculations also
show evidence of the hexagonal warping effect\cite{PhysRevLett.103.266801,PhysRevLett.105.076802} for $k_x\geq 0.08\pi/a$}. 

Now we discuss the thickness dependence in the bulk-Dirac energy gap measure and the spin polarization. We calculate the dimensionless measure of bulk-Dirac energy gap $r=\Delta_{DB}/\Delta_{BB}$ in the TM regime at the $\mathbf{k_{\parallel}}$ point where 
the Dirac surface state branch has slightly higher energy than the bottom of the conduction band $E_c$, as indicated in Fig.~\ref{DFT}(a). 
We find that the ratio $\Delta_{BB}(N_1)/\Delta_{BB}(N_2)$ is close to $(N_2/N_1)^2$ at the $k_{\parallel}$ point of interest as expected of finite-size-effect origin of the scale $\Delta_{BB}(N)$. Surprisingly, despite the small range of thickness accessible to the slab DFT calculation, the dimensionless measure of bulk-Dirac energy gap $r=\Delta_{DB}/\Delta_{BB}$ in Fig.~\ref{DFTNdep}(a) shows a significant increase upon an increase in the slab thickness. This is qualitatively consistent with observations from the effective model and hybridization effects in Sec.~\ref{sec:hyb}. 
On the other hand, the range of thickness in the present calculation appears to be too small to show any change in the 
 $\langle S_y \rangle$ as a function of slab thickness [Fig.~\ref{DFTNdep}(b)].

\section{Conclusion}
\label{sec:conclusion}

We combined a Fano-type hybridization model calculation with an ab-initio slab calculation to study the lowest order effects of surface-bulk interaction in topological insulators with a particular focus in the TM regime. We defined the TM regime of a topological insulator to be where the Dirac surface states and bulk states coexist and interact, yet the spin-winding is preserved albeit with a reduced spin-polarization magnitude. 
The hybridization model presented in Sec.~\ref{sec:hyb} captures the spin-polarization reduction of the Dirac states originating from the hybridization with bulk states. 
Given the metallic behavior of most TIs, and the experimental evidence of reduced spin polarization, our simple model offers a useful starting point for applications of TIs which need to take real materials in the TM regime into account.
Moreover, the hybridization-driven  bulk-Dirac energy gap explains why the Dirac branch shows up so well separated from bulk states in ARPES experiments. 
Note that this energy gap and the suppression of total spin magnitudes are both experimentally observed phenomena that cannot be accessed by the typical approach of 
 coupling a single ``surface layer" to a bulk electronic structure 
 to include surface states in semi-infinite systems as in Ref.~\onlinecite{Zhang2009}. 
We propose SARPES experiments for films of varying thickness to test our predictions for hybridization-driven suppression of spin polarization for further vindication of the model.

Promising future directions include DFT tools to study slightly thicker systems. This might reveal thickness dependence in spin polarization and compared to the results of the simple model. Also this would reveal more detailed knowledge of the magnitude and $\mathbf{k}_\parallel$-dependence of the hybridization strength $g$.
Preliminary DFT results show that $g(\mathbf{k}_\parallel)$ has a significant $\mathbf{k}_\parallel$-dependence. Another interesting direction will be to study consequences of the hybridization effect on transport properties. Many puzzling aspects of transport experiments\cite{PhysRevB.84.233101,PhysRevB.84.073109,PhysRevLett.109.116804,PhysRevB.83.165440} have been attributed to the presence of bulk states or surface-bulk interaction. There is growing theoretical interest on the transport properties of topological edge states in the presence of metallic bulk states\cite{PhysRevLett.107.176801,PhysRevLett.107.206602,PhysRevB.83.245432,PhysRevB.87.235114} as well. Our microscopic model offers a simple starting point to theoretically address effects of surface-bulk interaction on transport in 3D TIs. 

{\bf Acknowledgements.} We thank S. Oh for stimulating discussions. E.-A.K., Y.-T.H., and M.H.F.  were supported in part by NSF CAREER grant DMR-095582 and in part by the Cornell Center for Materials Research with funding from the NSF MRSEC program (DMR-1120296). K.P. was supported by NSF DMR-0804665, DMR-1206354, and San Diego Supercomputer Center (DMR-060009N). TLH was supported by  US DOE under QMN DE-FG02-07ER46453. 
\bibliography{my}

\end{document}